\documentclass[prl,aps,onecolumn,superscriptaddress]{revtex4}
\usepackage{amsfonts}

\usepackage{dcolumn}
\usepackage{bm}
\usepackage{tikz}
\usepackage{color}
\usepackage[colorlinks=true,citecolor=blue,urlcolor=blue]{hyperref}
\usepackage{amsmath,amsfonts,amssymb,times,natbib}
\usepackage[standard]{ntheorem}

\newcommand{\bra}[1]{\langle #1|}
\newcommand{\ket}[1]{|#1\rangle}

\begin{document}
\title{Conditional displacement interaction in ultrastrong-coupling regime}

\author{Gangcheng Wang\footnote{These authors contributed equally to the work}}
\affiliation{Center for Quantum Sciences and School of Physics, Northeast Normal University, Changchun 130024, China}

\author{Qingyong Wang\footnote{These authors contributed equally to the work}}
\affiliation{Center for Quantum Sciences and School of Physics, Northeast Normal University, Changchun 130024, China}

\author{Yimin Wang}
\affiliation{College of Communications Engineering, PLA University of Science and Technology, Nanjing 210007, China}
\affiliation{Beijing Computational Science Research Center, Beijing 100193, China}

\author{Jing-Ling Chen}
\affiliation{Theoretical Physics Division, Chern Institute of Mathematics, Nankai University, Tianjin 300071, China}
\affiliation{Centre for Quantum Technologies, National University of Singapore, 3 Science Drive 2, Singapore 117543}

\author{Kang Xue}
\affiliation{Center for Quantum Sciences and School of Physics, Northeast Normal University, Changchun 130024, China}

\author{Chunfeng Wu}
\email{chunfeng\_wu@sutd.edu.sg}
\affiliation{Pillar of Engineering Product Development, Singapore University of Technology and Design, 8 Somapah Road, Singapore 487372}

\date{\today}

\maketitle

\textbf{We investigate the realization of conditional displacement interaction in the transversal direction in ultrastrongly coupled circuit quantum electrodynamics by adjusting parameters of external magnetic fields. The special interaction is derived in the system of charge qubit(s) coupled to a LC resonator. We consolidate the implementation of quantum gates and the generation of superposed coherent states based on the transversal conditional displacement interaction numerically. The conditional displacement interaction in the ultrastrong coupling regime enhances quantum process to operate at the time scale of nanoseconds.}

\vspace{3mm}

The conditional displacement interaction depicts a quantum resonator conditionally displaced according to qubit(s)' states. The conditional interaction has played a prominent role in understanding the fundamentals in quantum physics~\cite{leggett2002,schwab2002,tian2016,monroe1996,haljan2005,yin2013,liu2005,liao2008,pender2015} and implementing quantum processing protocols~\cite{sorensen2000,garc2003,leibfried2003,feng2007,feng2009,blais2007,bill2015,zhu2003,zheng2004,kirchair2009,wang2002,didier2015}. Specifically, it has been widely utilized to generate superposed coherent states~\cite{schwab2002,tian2016,monroe1996,haljan2005,yin2013,liu2005,liao2008}, investigate quantum simulation~\cite{pender2015}, implement quantum gates~\cite{sorensen2000,garc2003,leibfried2003,feng2007,feng2009,blais2007,bill2015,zhu2003,zheng2004,kirchair2009,wang2002}, and achieve fast quantum nondemolition readout~\cite{didier2015}. It is worth mentioning that the conditional displacement interaction in the literatures has mainly been derived by applying the rotating-wave approximation (RWA) \cite{RWA} with respect to qubit-resonator coupling on the quantum Rabi model or the Jaynes-Cummings model~\cite{pender2015,sorensen2000,garc2003,leibfried2003,feng2007,feng2009,blais2007,bill2015,zhu2003,zheng2004,kirchair2009}. Of special note is the effective coupling strength with the RWA that is in general of order $10^{-3}\sim 10^{-2}\omega_r$, and would lead to quantum protocols operating at microseconds or less.

For the implementation of quantum gates, the operation time depends on the coupling strength between the qubits and the resonator. A stronger coupling is needed to speed up quantum gate operations, as well as to avoid the concomitant negative effects of decoherence during the evolution of the system. Actually ultrastrong coupling allows the realization of enhanced fast quantum gates operating at the time scale of sub-nanosecond~\cite{Garc2005,Monroe2010,romero2012}, and the deep strong coupling will decrease the time further by a factor of 10. In Refs.~\cite{gross2010, ana2009,gunter2009, todorov2010, scalari2012, pol2010, chen2016}, the ultrastrong coupling with superconducting qubits has been experimentally demonstrated. Very recently, the Semba group has experimentally achieved the deep strong coupling in circuit quantum electrodynamics (QED), where the rate between the coupling strength and the resonator frequency can reach $g/\omega_r \geq 1$ \cite{semba2016}. In the ultrastrong or deep strong coupling regime, the coupling strength $g$ is comparable to or greater than the resonator frequency and hence the RWA is no longer applicable in respect of the qubit-resonator coupling. Therefore, it is not possible to use the RWA with respect to the coupling strength to obtain the conditional displacement, but to use the direct quantization of the system interaction. In 2012, Ref.~\cite{romero2012} discussed a special design of superconducting flux qubits in order to realize the interaction in the longitudinal direction in the ultrastrong coupling regime. However, the complicated design of flux qubits makes it difficult to experimentally realize the ultrastrong interaction.

In this work, we explore the conditional displacement interaction in the transversal direction in the ultrastrong coupling regime by resorting to parametric modulation of external magnetic fields. The special interaction is derived in the superconducting circuits of charge qubit(s) coupled to a LC resonator. The ultrastrong conditional displacement interaction enhances quantum process to operate at the time scale of nanoseconds in the transversal direction. Our scheme paves a promising way to speed up quantum information processing according to the conditional displacement  interaction. We then explore the implementation of quantum gates and the generation of Schr\"{o}dinger cat states based on the conditional displacement interaction with numerical results.

\vspace{8pt}
\noindent{\bf Results}

\begin{figure}[tbp]
\begin{center}
\includegraphics[width=0.4\textwidth]{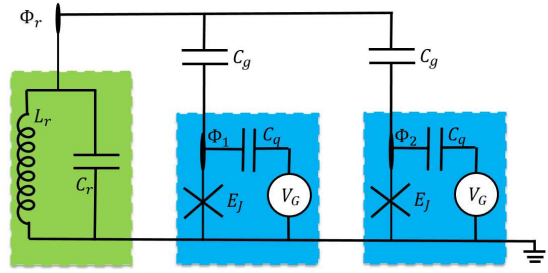}
\caption{(Color online) Illustration of $2$ charge qubits coupled to a LC resonator.}
\label{fig1}
\end{center}
\end{figure}

\noindent{\bf The system of charge qubit(s) coupled to a LC resonator.}
We consider a system of two charge qubits coupled to a LC resonator as shown in Fig.~\ref{fig1}. The system Hamiltonian is of the form ($\hbar = 1$)~\cite{rabl16},
\begin{eqnarray}\label{eq2}
H &=& \sum_{m=1}^{2}\frac{\omega_{q}^{m}}{2}\sigma_{z}^{m}+ \omega_{r}a^{\dag}a+\sum_{m=1}^{2}g_m(a^{\dag}+a)\sigma_{x}^{m}+\sum_{m\neq n=1}^{2}D_{mn}\sigma_{x}^{m}\sigma_{x}^{n},\nonumber\\
\end{eqnarray}
where $\omega_{q}^{m}$ is the energy splitting of $m$-th qubit, $\sigma_{\alpha}^{m}$ is the $\alpha$ component of the $m$-th Pauli matrix, $\omega_{r}$ is the frequency of the resonator, $a$ ($a^{\dag}$) is the annihilation (creation) operator, $g_m$ is the coupling constant, and $D_{mn}=\frac{g_mg_n}{\omega_r}$. For the sake of simplicity, we assume that $\omega_{q}^{m}=\omega_q$ and $g_m=g$, then we have $D_{mn}=D$.

Apply external magnetic fields defined in the following form on qubits such that it is possible to do parametric modulation~\cite{xue2015, xue2015_2, eckardt2015, strand2013},
\begin{eqnarray}\label{pm}
  \omega_{q}^{m}(t) = \omega_{q} + \varepsilon_{m}\sin(\omega_{m}t-\phi_{m})   \, .
\end{eqnarray}
Move to the rotating frame defined by time-dependent transformation $\mathcal{U}(t) = \mathcal{U}_{1}(t)\mathcal{U}_{2}(t)$ with
\begin{subequations}\label{eq:7}
\begin{align}
 &\mathcal{U}_{1}(t) =  \exp\left[-i\left(\sum_{m}\frac{\omega_{q}}{2}\sigma_{z}^{m} + \omega_{r}a^{\dag}a\right)t\right]\\
&\mathcal{U}_{2}(t) = \exp\left[i\sum_{m}\frac{\alpha_{m}}{2}\cos(\omega_{m}t-\phi_{m})\sigma_{z}^{m}\right],
   \end{align}
\end{subequations}
where $\alpha_{m}=\varepsilon_{m}/\omega_{m}$. The transformed Hamiltonian $\tilde{H}_{1} = \mathcal{U}^{\dag}(t)H\mathcal{U}(t)-i\mathcal{U}^{\dag}(t)\partial_{t}\mathcal{U}(t)$ with $\phi_{m}=\phi$ and $\omega_{m}=\omega$ reads
\begin{eqnarray}
\label{eq:H}
  \tilde{H}_1(t) &=& g\sum_{m=1}^2\left[ a^{\dag}\sigma_{-}^{m}e^{i\delta_{-}t}e^{i\alpha_{m}\cos(\omega t-\phi)} + h.c.\right] \nonumber\\
        &+& g\sum_{m=1}^2\left[ a^{\dag}\sigma_{+}^{m}e^{i\delta_{+}t}e^{-i\alpha_{m}\cos(\omega t-\phi)} + h.c.\right]\nonumber\\
        &+& D\sum_{m\neq n=1}^{2}\left[ \sigma_{+}^{m}\sigma_{+}^{n}e^{i2\omega_q t}e^{-i(\alpha_{m}+\alpha_{n})\cos(\omega t-\phi)} + h.c.\right]\nonumber\\
        &+& D\sum_{m\neq n=1}^{2}\left[ \sigma_{+}^{m}\sigma_{-}^{n}e^{-i(\alpha_{m}-\alpha_{n})\cos(\omega t-\phi)} + h.c.\right],
\end{eqnarray}
where $\delta_{\pm} = \omega_{r}\pm \omega_{q}$. By applying the Jacobi-Anger expansion~\cite{colton1998},
\begin{eqnarray}
  \exp\left(i\alpha_{m}\cos(\omega t-\phi )\right) = \sum_{l=-\infty}^{+\infty} i^{l}J_{l}(\alpha_{m})\exp(il(\omega t-\phi))   \, ,
\end{eqnarray}
where $J_{l}(\alpha_m)$ is the Bessel function of first kind. If we set $\omega_{q}=\omega=\eta\omega_{r}$ with $\eta > 2$, the lowest oscillation frequency is $\omega_{r}$. Ignoring all the higher-order terms, we can obtain the following effective Hamiltonian (See Methods),
 \begin{eqnarray}
\label{eq:Htilde_1}
  \tilde{H}_1^{\prime} &=& \sum_{m=1}^2 g^m_{\rm eff}\left(a^{\dag}e^{i\omega_{r}t}+a e^{-i\omega_{r}t}\right)\sigma_{x}^{m},
\end{eqnarray}
where $g^m_{\rm eff}=gJ_1(\alpha_m)$, and we select $\alpha_1-\alpha_2=2.40483$ and $\phi=\pi/2$. This is just the desired conditional displacement interaction in the transversal direction for two qubits. In the case of one qubit, the conditional displacement interaction is also possible with the condition that $\eta > 2$ and $\phi=\pi/2$. In particular, the effective coupling strength is adjustable dependent on the value of $J_1(\alpha_m)$ with $m=1$. While for two qubits, the $g_{\rm eff}^m$ is also tunable and it is possible to realize different $g_{\rm eff}^m$ for individual qubits or $g_{\rm eff}^1=-g_{\rm eff}^2$, as long as the condition $\alpha_1-\alpha_2=2.40483$ is fulfilled. The effective coupling strength of the conditional interaction reaches its maximum value of $gJ_1(1.832)\approx 0.582g$ when $\alpha_m=1.832$. When $\alpha_1=-\alpha_2=1.20242$, we find $g^1_{\rm eff}=-g^2_{\rm eff}=gJ_1(1.20242)\approx 0.499g$.

\begin{figure}[tbp]
\begin{center}
\includegraphics[width=0.4\textwidth]{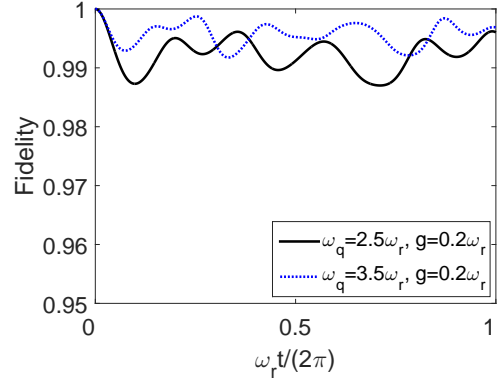}
\caption{The fidelity $\mathcal{F}_1$ varies as function of the evolution time $t$ for different parameters, where black solid curve is for $\omega_{q}=2.5\omega_{r}$ and $g=0.2\omega_{r}$, and blue dotted curve is for $\omega_{q}=3.5\omega_{r}$ and $g=0.2\omega_{r}$.}
\label{fig2}
\end{center}
\end{figure}

In the following, we check the validity of the approximation in a numerical way by taking $N=2$ as an example. Choose $\ket{\psi_0}=\ket{gg0_c}$ as the initial state, where $\ket{g}$ is the ground state of the charge qubit and $\ket{0_c}$ is the vacuum state of the LC resonator, and denote $\ket{\psi_{f}(t)}$ and $\ket{\psi^{\prime}_{f}(t)}$ as the evolution states governed by the original Hamiltonian with external magnetic fields and the effective Hamiltonian (\ref{eq:Htilde_1}), correspondingly. Let $\mathcal{F}_1(t)=|\bra{\psi_{f}(t)}\psi^{\prime}_{f}(t)\rangle|^{2}$ be the fidelity for the states $\ket{\psi_{f}(t)}$ and $\ket{\psi^{\prime}_{f}(t)}$. We numerically find $\mathcal{F}_1$ at $T=2\pi/\omega_r$ by choosing $\alpha_1=-\alpha_2=1.20242$ and the results are shown in Fig. \ref{fig2}, where black solid curve is for $\omega_{q}=2.5\omega_{r}$ ({\it i.e.} $\eta = 2.5$), $g=0.2\omega_{r}$, and blue dotted curve is for $\omega_{q}=3.5\omega_{r}$ ({\it i.e.} $\eta = 3.5$), $g=0.2\omega_{r}$. It can be inferred from the numerical results that the effective Hamiltonian (\ref{eq:Htilde_1}) is approximating the original time-dependent Hamiltonian very well. The larger the value of $\eta$, the more excellent approximation can be obtained given a fixed value of $g$.

\begin{figure}[tbp]
\begin{center}
\includegraphics[width=0.4\textwidth]{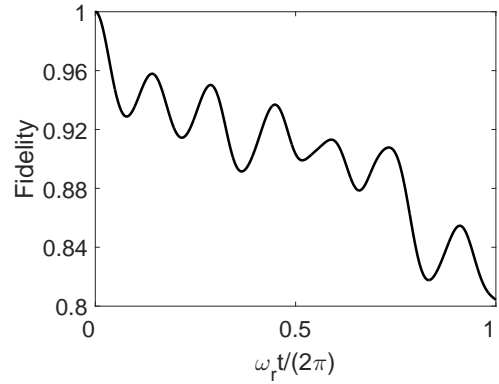}
\caption{The fidelity $\mathcal{F}_1$ varies as function of the evolution time $t$ for the following parameters $\omega_{q}=3\omega_{r}$ and $g=0.5\omega_{r}$.}
\label{fig3}
\end{center}
\end{figure}

Moreover, when $g$ is increasing, larger value of $\eta$ is required to achieve very good approximation according to the conditions mentioned. However, in practical experiments, the required strong driving may not be achievable. As a result, the fidelity will be decreased with currently achievable parameters. We show the numerical results in Fig. \ref{fig3}, where the parameters are chosen as $\omega_{q}=3\omega_{r}$ ({\it i.e.} $\eta = 3$), $g=0.5\omega_{r}$. As demonstrated in Fig. \ref{fig3}, with comparatively small external driving strength, the fidelity is affected largely when $g=0.5\omega_{r}$. This tells us in the case that the external driving is not comparatively large, our scheme performs well when $g$ is around $0.2\omega_{r}$, but becomes less and less desirable when $g$ is increasing.

\noindent{\bf Quantum gate with the conditional interaction in ultrastrong coupling regime.}
We next study the applications of the transversal conditional displacement interaction in implementing two-qubit gates. We choose $g^1_{\rm eff}=-g^2_{\rm eff}=g_{\rm eff}$ for the sake of simplicity in the following discussions. In the case of two qubits, the evolution operator obtained from the transversal interaction is~\cite{wang2002,zhu2003,zheng2004,kirchair2009}
\begin{eqnarray}
\label{eq:U}
  U(t) &=& D\left[\beta(t)J_{x}\right]\exp\left(i\Phi(t)J_{x}^{2}\right),
\end{eqnarray}
where $J_{x}=\sigma_{x}^{1}-\sigma_{x}^{2}$, $\beta(t)=(g_{\rm eff}/\omega_{r})(1-e^{i\omega_{r}t})$ and $\Phi(t)=(g_{\rm eff}/\omega_{r})^{2}(\omega_{r}t-\sin(\omega_{r}t))$, and $D(\beta)=e^{\beta a^{\dag} - \beta^{*}a}$. Let the system evolve for a time period of $T = 2\pi/\omega_{r}$, we obtain $\beta(T)=0$ and $\Phi(T)=2\pi(g_{\rm eff}/\omega_{r})^{2}$. Then evolution operator can be recast as follows~\cite{wang2002,zhu2003,zheng2004,kirchair2009},
\begin{eqnarray}
\label{eq:U_1}
  U(T) &=& \exp\left(i2\pi(g_{\rm eff}/\omega_{r})^{2}J_{x}^{2}\right).
\end{eqnarray}
We thus have a tunable quantum phase gate and it can be rewritten as $U(T) = e^{i\theta}e^{-i\theta\sigma_{x}^{1}\sigma_{x}^{2}}$ with $\theta = 4\pi(g_{\rm eff}/\omega_{r})^{2}$. Express the evolution operator in the basis of $\{\ket{ee},\ket{eg}, \ket{ge}, \ket{gg}\}$, we obtain
\begin{eqnarray}
\label{eq:U_matrix}
  U(T) &=& e^{i\theta}\left(
  \begin{matrix}
  \cos\theta & 0 & 0 & -i\sin\theta \\
  0 & \cos\theta & -i\sin\theta & 0 \\
  0 & -i\sin\theta & \cos\theta & 0 \\
  -i\sin\theta & 0 & 0 & \cos\theta
\end{matrix}\right).
\end{eqnarray}
The evolution operator $U(T)$ represents non-trivial two-qubit gates when $\theta \neq n\pi\; (n= 0,\pm 1,\pm 2,\cdots)$. Specifically when $\theta = \pi/4$ ({\it i.e.}, $g_{\rm eff} = 0.25\omega_{r}$), $U(T)$ is locally equivalent to the control-NOT (CNOT) gate \cite{rezak2004}.

We study the performance of the quantum gate in the following by generating $N = 50$ random two-qubit states of the form $\ket{\psi(a_i)}=\frac{1}{\sqrt{\sum_{i=1}^4|a_i|^2}}\big(a_1\ket{ee}+a_2\ket{eg}+a_3\ket{ge}+a_4\ket{gg}\big)\otimes \ket{0_c}$ as initial states, where $i=1,2,3,4$.
We numerically calculate average fidelity defined by $\overline{F}_G = \overline{|\bra{\psi_{1,{\rm ideal}}}U'(T)\ket{\psi(a_i)}|^2}$, where $\ket{\psi_{1,{\rm ideal}}}=U(T)\ket{\psi(a_i)}$ and $U'(T)$ is the evolution operator based on Hamiltonian (\ref{eq2}) with external magnetic fields. The average fidelity is found to be $\overline{F}_G = 0.9948$ when $\omega_{q}=3\omega_{r}$ ({\it i.e.} $\eta = 3$), $g=0.2\omega_{r}$ and $\alpha_1=-\alpha_2=1.20242$. The numerical results show the excellent performance of our scheme of achieving the two-qubit quantum gate with $\theta=\frac{\pi}{25}$ in the ultrastrongly coupled circuit system. Moreover, we also explore the performance of our scheme to achieve the quantum gate with $\theta=\frac{\pi}{4}$ by choosing $\omega_q=3\omega_{r}$ ({\it i.e.} $\eta = 3$), $g=0.5\omega_{r}$ and $\alpha_1=-\alpha_2=1.20242$, and the fidelity is $0.7844$. Further improvement in the gate fidelity is dependent on the development of experimental techniques to realize very strong external driving. Therefore we achieve to implement two-qubit quantum phase gate at the time scale of nanoseconds depending on the value of $\omega_r$. The scheme possesses the merit of enhanced fast quantum operation to avoid the detrimental effect of decoherence. Even thought the fidelity of the gate with $\theta=\frac{\pi}{4}$ is not very good with current experimental techniques, our scheme is still a noticeable improvement in achieving fast universal quantum computation in which any type of non-trivial two-qubit gate is desired.

\noindent{\bf Schr\"{o}dinger cat states with the conditional interaction in ultrastrong coupling regime.}
The conditional interaction is also crucial in creating superposed coherent states and hence exploring the superposition rule. To get clear evidence of the quantum superposition,  the displacement of the resonator should be maintained rather than destroyed by decoherence. Thus enhanced fast generation of the superposed coherent states is demanded to avoid the negative effect of decoherence. We then investigate the creation of the superposed coherent states with the ultrastrong transversal interaction.

Take the case of one qubit coupled to resonator as an example, the evolution operator is represented by Eq.~(\ref{eq:U}) by replacing $J_x$ by $\sigma_x$. Let $\ket{\Psi_1(0)} = \ket{g}\otimes \ket{0_c}$ be initial state, and we obtain the final state at time $t$ as $\ket{\Psi_1(t)} = \frac{e^{i\Phi(t)}}{\sqrt{2}}\left(\ket{+}\otimes \ket{\beta(t)_c}-\ket{-} \otimes \ket{-\beta(t)_c}\right)$, where $\ket{\pm}=\frac{1}{\sqrt{2}}(\ket{e}\pm\ket{g})$ and the coherent states $\ket{\pm\beta(t)_{c}}=D[\pm\beta(t)]\ket{0_{c}}$ with coherent-state amplitude $\pm\beta(t)_{c}=\pm(g_{\rm eff}/\omega_{r})(1-e^{i\omega_{r}t})$. Obviously, the spin states $\ket{+}$ and $\ket{-}$ undergo different displacement $\beta(t)_c$ and $-\beta(t)_c$, respectively. In the basis of $\{\ket{g},\ket{e}\}$, the final state can be rewritten as~\cite{tian2016,monroe1996,haljan2005,yin2013,liu2005,liao2008,ge2015,david1996,pawlowski2016},
\begin{eqnarray}
\label{cat_1bit}
\ket{\Psi_1(t)} &=& \frac{e^{i\Phi(t)}}{2}\left(N_{0}^{-1}\ket{g}\otimes\ket{\beta_{0}(t)_c}+N_{1}^{-1}\ket{e}\otimes\ket{\beta_{1}(t)_c}\right)
\end{eqnarray}
where $N_{l}=\left[2(1+(-1)^l \exp(-2|\beta(t)|^{2}))\right]^{-1/2}$ and $\ket{\beta_{l}(t)_c}=N_{l}(\ket{\beta(t)_c}+(-1)^l\ket{-\beta(t)_c})$ with $l=0,1$. The superpositions of coherent states $\ket{\beta_{0}(t)_c}$ and $\ket{\beta_{1}(t)_c}$ are the so-called even and odd Schr\"odinger states.

Performing projective measurement in the qubit basis, superposed coherent states $\ket{\beta_{l}(t)_c}$ can be extracted with probability of $\frac{1}{2}\bigg[1+(-1)^l \exp\big(-2|\beta(t)|^{2}\big)\bigg]$, respectively. The magnitude of the displacement is dependent on evolution time $t$. When $t_0=\pi/\omega_r$, we obtain maximum magnitude of $\beta(t_0)_c=2g_{\rm eff}/\omega_r$. In the case of single charge qubit coupled to the LC resonator, the original time-dependent Hamiltonian is depicted by (\ref{eq2}) with external magnetic fields. We find the fidelity between ideal state (\ref{cat_1bit}) and the final state solved numerically from the original time-dependent Hamiltonian. The parameters are chosen as $\omega_{q}=3\omega_{r}$, $g=0.2\omega_{r}$, and $\alpha_{1}=1.832$ and the realized displacement is $0.233$. According to our numerical calculations, state (\ref{cat_1bit}) and hence the superposed coherent states can be created with a fidelity of $0.9978$. The amplitude of the displacement can be enhanced via multi-step system evolution. Specifically, let the initial state be $\ket{g0_{c}}$, and after the time period $t_{0}$, we obtain state $\ket{\Psi_{1}}$. Acting the evolution operator on the state $\ket{\Psi_{1}}$ for another $t_0$, we obtain state $\ket{\Psi_{2}}=\frac{e^{i2\Phi(t_0)}}{\sqrt{2}}\left(\ket{+}\otimes \ket{2\beta(t_0)_c}-\ket{-} \otimes \ket{-2\beta(t_0)_c}\right)$. Therefore the displacement amplitude is $2\beta(t_0)_c=4g_{\rm eff}/\omega_{r}$ after two steps of evolution with a total evolution time $2\pi/\omega_r$. The corresponding fidelity of generating $\ket{\Psi_{2}}$ is $0.9902$. Following the multi-step evolution, the amplitude of displacement can be further enhanced.

Based on current experimental techniques of superconducting circuits, it may be possible to realize our scheme in the superconducting system. In our scheme, it is desired that the qubits are ultrastrongly coupled to the resonator with strong driving external magnetic fields. As an example, the required parameters may be $\omega_r\sim2\pi\times 2$ GHz, $\omega_q=\omega\sim2\pi\times 6$ GHz and $\varepsilon\sim 2\pi\times 7.2$ GHz. The reported strong driving of superconducting qubits in experiments is roughly $2\pi \times 5$ GHz~\cite{lupascu2015}, a little below the required driving strength. As for the required ultrastrong qubit-resonator coupling strength, there are noticeable achievements in the known literature. In Refs.~\cite{gross2010,pol2010}, the authors have experimentally achieved the ultrastrong coupling with flux qubits coupled to a transmission line resonator, and the coupling strength is around $0.1\omega_r$~\cite{gross2010,pol2010}. For the system of superconducting qubits coupled to a LC resonator, the Semba group has extended the qubit-resonator coupling to the deep strong coupling regime experimentally~\cite{semba2016}. It can be inferred that our scheme to achieve the transversal interaction is possibly achievable with the state-of-art techniques.

\vspace{8pt}
\noindent{\bf Methods}

By applying the Jacobi-Anger expansion~\cite{colton1998}, we find
\begin{eqnarray}
\label{eq:H1}
  \tilde{H}_1(t) &=& g\sum_{m=1}^2\left[ a^{\dag}\sigma_{-}^{m}\sum_{l=-\infty}^{+\infty} i^{l}J_{l}(\alpha_{m})e^{i\delta_{-}t}e^{il(\omega t-\phi)} + h.c.\right] \nonumber\\
        &+& g\sum_{m=1}^2\left[ a^{\dag}\sigma_{+}^{m}\sum_{l=-\infty}^{+\infty} (-i)^{l}J_{l}(\alpha_{m})e^{i\delta_{+}t}e^{-il(\omega t-\phi)} + h.c.\right]\nonumber\\
        &+& D\sum_{m\neq n=1}^{2}\left[ \sigma_{+}^{m}\sigma_{+}^{n}\sum_{l=-\infty}^{+\infty} (-i)^{l}J_{l}(\alpha_{m}+\alpha_{n})e^{i2\omega_q t}e^{-il(\omega t-\phi)} + h.c.\right]\nonumber\\
        &+& D\sum_{m\neq n=1}^{2}\left[ \sigma_{+}^{m}\sigma_{-}^{n}\sum_{l=-\infty}^{+\infty} (-i)^{l}J_{l}(\alpha_{m}-\alpha_{n})e^{-il(\omega t-\phi)} + h.c.\right].
\end{eqnarray}
If we set $\omega_{q}=\omega=\eta\omega_{r}$ with $\eta > 2$, the lowest oscillation frequency is $\omega_{r}$ and hence many higher-order terms can be neglected. Take the first line in Eq. (\ref{eq:H1}) as an example, we have
\begin{eqnarray}
\tilde{H}_1^{(1)}(t)=g\sum_{m=1}^2\left[ a^{\dag}\sigma_{-}^{m}\sum_{l=-\infty}^{+\infty} i^{l}J_{l}(\alpha_{m})e^{i\delta_{-}t}e^{il(\omega t-\phi)} + h.c.\right].
\end{eqnarray}
When $l=0$, the phases reduce to $e^{\pm i\delta_{-}t}=e^{\pm i(\omega_{r}-\omega_{q})t}$. In the case that $\omega_q > 2\omega_r$ with proper choice of parameters, it is possible to make $|\delta_{-}|\gg g|J_{0}(\alpha_{m})|$ and hence it gives a higher-order oscillating term which can be omitted. When $l=1$, the phases are combined as $e^{\pm i(\delta_{-}+\omega) t\mp i\phi}=e^{\pm i\omega_{r}t\mp i\phi}$ since $\omega_q=\omega$. The system is in the ultra-strong coupling regime and when $\omega_{r}$ is not much greater than $g|J_{1}(\alpha_{m})|$, it cannot be neglected according to the RWA. When $l=-1$, the phases are revised as $e^{\pm i(\delta_{-}-\omega) t\pm i\phi}=e^{\pm i(\omega_{r}-2\omega_{q})t\pm i\phi}$. By properly choosing parameters, we have $|\omega_{r}-2\omega_{q}|\gg g|J_{-1}(\alpha_{m})|$, and this tells us that the term can be neglected too. Following similar reasoning, it can be found that all the terms in $\tilde{H}_1^{(1)}(t)$ with $l\neq 1$ are higher-order oscillating terms. While for the second line in Eq. (\ref{eq:H1}), all the terms with $l\neq 1$ are also higher-order oscillating terms based on appropriate parameters. Next let us look the third line in Eq. (\ref{eq:H1}), it is found that most of the terms are higher-order oscillating terms except $l=2$. In this case the coefficient is $D|J_{2}(\alpha_{m}+\alpha_{n})|$ which is much smaller than $g$ even when $D$ is comparable with $g$, because $J_{2}(\alpha_{m}+\alpha_{n})$ can be chosen to be much smaller than 1 and therefore the term can be omitted. Lastly for the fourth line in Eq. (\ref{eq:H1}), the only non-higher-order oscillating term is given when $l=0$ and the coefficient is $DJ_{0}(\alpha_{m}-\alpha_{n})$ which can be set to be 0 when $\alpha_{m}-\alpha_{n}=2.40483$. Finally ignoring all the higher-order terms, we can obtain the effective Hamiltonian (\ref{eq:Htilde_1}).

\vspace{8pt}
\noindent{\bf Discussion}

In this work, we have proposed to achieve conditional displacement interaction of qubits and resonator in the transversal direction in ultrastrongly coupled superconducting system. The system is composed of charge qubits coupled to a LC resonator. By adjusting external magnetic fields in the system, the conditional displacement interaction is obtained for one or two qubits. The effective coupling strength is controllable dependent on the value of $\alpha_m$, and it is possible to realize different $g_{\rm eff}^m$ for individual qubits or $g_{\rm eff}^1=-g_{\rm eff}^2$ for two qubits. The effective coupling strength of the conditional interaction reaches its maximum value of $0.582g$ when $\alpha_m=1.832$ and $g_{\rm eff}^1=-g_{\rm eff}^2=0.499g$ when $\alpha_1=-\alpha_2=1.20242$. We have revealed numerically that the effective Hamiltonian is very close to the original Hamiltonian with external magnetic fields. Based on the transversal interaction, the implementation of two-qubit quantum phase gates and the generation of superposed coherent states are achievable at the time scale of nanosecond. Finally we have provided some discussions about the feasibility of the conditional displacement interaction according to the current experimental techniques.

\vspace{8pt}
{\bf Supplementary Information} is linked to the online version of the paper at www.nature.com/nature.
\vspace{2pt}

{\bf Acknowledgements}

The work is supported by the NSF of China (Grant Nos. 11405026 and 11575042). Y.M.W. is partly supported by the NSF of China (Grant No.\ 11404407), the NSF of Jiangsu (Grant No.\ BK20140072) and China Postdoctoral Science Foundation (Grant Nos.\ 2015M580965 and 2016T90028). J.L.C. is supported by the NSF of China (Grant No. 11475089).

\vspace{2pt}

{\bf Author contributions}

C.W. initiated the idea. All authors developed the scheme and wrote the main manuscript text.

\vspace{2pt}

{\bf Additional information}

Competing financial interests: The authors declare no competing financial interests.

Correspondence and requests for materials should be addressed to C.W. (chunfeng\_wu@sutd.edu.sg).

\newpage

\end{document}